# Controllable gas sensitivity of Mg/CuO nanocomposite films measured in methanol vapor

C.A. Samarahewa

Department of Physics, University of Peradeniya, Peradeniya, Sri Lanka

**Abstract**

The gas sensitivity of Mg/CuO nanocomposite films, characterized by varying mass ratios of Mg:CuO, was assessed under exposure to 1000 ppm of methanol vapor. Films were fabricated by the doctor blade method on conductive and nonconductive glass substrates. Structural and optical analyses were conducted using XRD and UV-Visible spectrums. The XRD patterns facilitated the estimation of crystallite size, dislocation density and strain. The optical band gap of the samples was determined from UV-Visible spectrums. Despite variations in crystallite size, dislocation density and strain in response to changing Mg concentrations in nanocomposites, no discernible shift in the band gap was observed. The mass percentage of Mg in nanocomposite was incrementally altered from 10% to 20% in steps of 5%. Due to the adsorption of methanol vapor, the resistivity of the sample decreased significantly. Gas sensitivity exhibited variance ranging from 3.79 for pure CuO to 1.23 for nanocomposite with 20%Mg. The sample with 10%Mg quickly responded to methanol vapor compared to pure CuO and other nanocomposites.

## 1 .Introduction:

Two copper based oxide materials are CuO (copper oxide) and $Cu_2O$ (cuprous oxide). CuO is a material with a monoclinic structure and band gap of 1.2 to 1.9 eV. The band gap of $Cu_2O$ is slightly higher than that of CuO. CuO finds potential applications in gas sensors, photovoltaic, supercapacitors, near infrared sensors, catalysis and magnetic memory devices. Mg is a material with a hexagonal closed packed structure and very small band gap. Thin films of CuO have been synthesized by many techniques for various applications. Thin films of CuO have been deposited by precipitation technique. Elongated spherical nanoparticles of CuO with antiferromagnetic properties have been obtained [1]. Dip coating method has been employed to fabricate CuO films. The thickness of the films has been varied by changing the number of coats [2]. In addition, CuO films have been prepared by the thermal evaporation technique. The optimum photocatalytic properties of these CuO films have been observed for the films annealed at 400 0C [3]. The dip coating technique has been applied to synthesize CuO and $Cu_2O$ films. Phases of CuO and $Cu_2O$ have been prepared at the temperatures of 230 and 260 $^0$C, respectively [4]. Furthermore, copper oxide films have been synthesized on Si (111) substrates using spray pyrolysis. The gas sensing properties of these films have been investigated at different operating temperatures [5].

Metal oxides such as CuO, $Cu_2O$, ZnO, $WO_3$ and $Fe_2O_3$ are mostly used in gas sensing applications. $WO_3$ films deposited on micropatterned gold electrodes and annealed at 500 $^0$C in air for 2 hours have been employed to detect methane and nitric oxide gases. The gas sensitivity depends on the thickness of $WO_3$ films and the operating temperature. The response and recovery of gas sensors were faster for thinner films. The grain boundary control method has been applied to explain the origin of the gas sensitivity [6]. ZnO films prepared using the doctor blade method has been employed to detect ethanol, methanol, acetone vapors and $CO_2$ gas. The lowest recovery time (6 min) has been measured in acetone vapor compared to ethanol, methanol vapors and $CO_2$ gas. The lowest response time (4 min) of these ZnO films has been measured in $CO_2$ gas compared to other vapors. The higher gas sensitivity (65.5%) was measured in acetone [7]. Copper oxide films have been deposited by DC reactive sputtering. A pure Cu target with a sufficient amount of oxygen in a vacuum chamber has been employed to form the phase of CuO. The effect of sputtering conditions on the gas sensitivity of CuO films measured in $CO_2$ gas has been investigated [8]. In addition, the effect of the PEG binder on the gas sensitivity of $\alpha$-$Fe_2O_3$ films synthesized using the doctor blade and the spin coating methods measured in 1000 ppm of $CO_2$ gas has been studied. The addition of the PEG binder has increased the gas sensitivity by 15.44%, since PEG binder has decreased the activation energy of $\alpha$-$Fe_2O_3$. However, the response and recovery times slightly changed due to the addition of PEG binder [9]. The gas sensitivity of $\alpha$-$Fe_2O_3$ films has been measured in acetone, ethanol, methanol vapors and $CO_2$ gas at different operating temperatures from 28 to 200 $^0$C. The highest gas sensitivity has been found at 170 $^0$C operating temperature. At the room temperature, the highest gas sensitivity was measured for $CO_2$ compared to acetone, ethanol and methanol vapors [10]. The gas sensitivity of ferric oxide films has been enhanced by doping Mn nanoparticles. The mass ratio of Mn in $\alpha$- $Fe_2O_3$:Mn has been varied from 3 to 10%. Thin films with 6% Mn doping concentration have indicated the highest gas sensitivity of 70.1% at the room temperature as measured in 1000 ppm of $CO_2$ [11]. Furthermore, the gas sensitivity of ferric oxide films has been improved by doping commercially available activated carbon nanoparticles. The highest gas sensitivity (50.2%) has been measured for 2% doping concentration of activated carbon in 1000 ppm in $CO_2$ [12].

Thin films of CuO have been synthesized by the sol gel method with different precursors such as copper acetate and 2-methoxy-ethanol. A decrease of the band gap from 1.67 to 1.56 eV has been observed with an increasing number of layers in the film [13]. CuO films with different thicknesses have been fabricated by rf magnetron sputtering. These films were found to be applicable in the degradation of methylene blue from wastewater [14]. In addition, CuO films have been deposited by reactive magnetron sputtering for different oxygen flow rates for solar energy applications. Films with uniform surfaces were observed at lower oxygen flow rates [15]. A successive ionic layer adsorption and reaction method have been employed to prepare CuO films. Films have indicated better crystallization as the number of deposition cycle was increased from 30 to 50 cycles [16]. Microwave activated reactive sputtering method was used to deposit $Cu_2O$, CuO and $Cu_4O_3$ films. According to XRD patterns, $Cu_2O$ and $Cu_4O_3$ phases were detected at oxygen flow rates from 11 to 16 sccm [17]. Furthermore, Spray pyrolysis method has been employed to fabricate CuO films to detect acetone. The gas sensitivity, response time and recovery time of these films in acetone were found to be 33%, 160 s and 360 s, respectively [18]. Response and recovery times of CuO films in low ppm of ethanol vapor were 52 and 42 s. As the ethanol vapor concentration was increased from 10 to 500 ppm, the gas sensitivity has increased from 1.3 to 3.3 [19].

In this manuscript, we present structural, optical and the gas sensing properties of Mg/CuO nanocomposite thin films. The gas sensitivity, response time and recovery time vary with the composition of nanocomposite.

## 2. Experimental:

Polyethylene glycol (PEG) was used as the binder. 0.06 g of PEG powder mixed with 8 ml of distilled water was stirred at 50 $^0$C at 300 rpm for 15 min. CuO nanoparticles manufactured by Johnson Matthey materials technology was used to prepare thin films. 1.5 g of CuO powder or CuO:Mg nanocomposite was mixed with 5 ml of PEG, and stirred at 50 $^0$C at 600 rpm for 15 min. The amount of Mg in CuO:Mg nanocomposite was varied from 10 to 20% in steps of 5%. The synthesized samples were subsequently annealed at 100 $^0$C in air for 1 hour. The area of the sample was 2.55 cm$^2$.

The structural properties of films deposited on non-conductive amorphous glass substrates were determined using a Rigaku Ultima IV X-Ray diffractometer with Cu-Kα (λ=1.5406 Å) radiation. The optical band gap of films synthesized on non-conductive amorphous glass substrates was found by means of a a Shimadzu 1800 UV-Vis spectrophotometer.

For the gas sensitivity measurements, the sample was fabricated on a conductive glass plates. The center part of conductive layer was removed so that the electric current flows through the sample. A circuit consists of the sample and a standard resistor connected in series with a 5V DC power supply was used to measure the gas sensitivity. The value of the standard resistor was equal to the resistance of the sample. Gold coated Cu wires were employed in order to provide better electrical contacts. The sample was placed inside a sealed chamber. The voltage across the standard resistor was measured. After introducing the methanol vapor into the chamber, the voltage across the standard resistor increased. While removing the gas in the chamber, the voltage across the standard resistor decreased.

## 3. Results and discussion:

Figure 1 and 2 show XRD patterns of Mg/CuO nanocomposites and pure Mg samples, respectively. The miller indices of peaks are marked in the XRD pattern of the CuO sample. All the peaks in figure 2 match with XRD peaks of pure Mg [20]. Only the peaks of pure CuO are evident in all nanocomposite samples. At 10%Mg doped nanocomposite, Mg atoms may occupy the vacant sites in CuO lattice. As a result, the peaks of Mg are not visible in the XRD pattern of nanocomposite with 10%Mg. However, the doping is not possible at higher concentrations of Mg such as 15% and 20%. In these higher concentrations of Mg, the phase of Mg may have filled the grain boundaries. As the Mg concentration is increased, the XRD peaks of CuO diminish in nanocomposites. For the sample with 20% Mg, the intensities of XRD peaks of CuO are really diminutive.

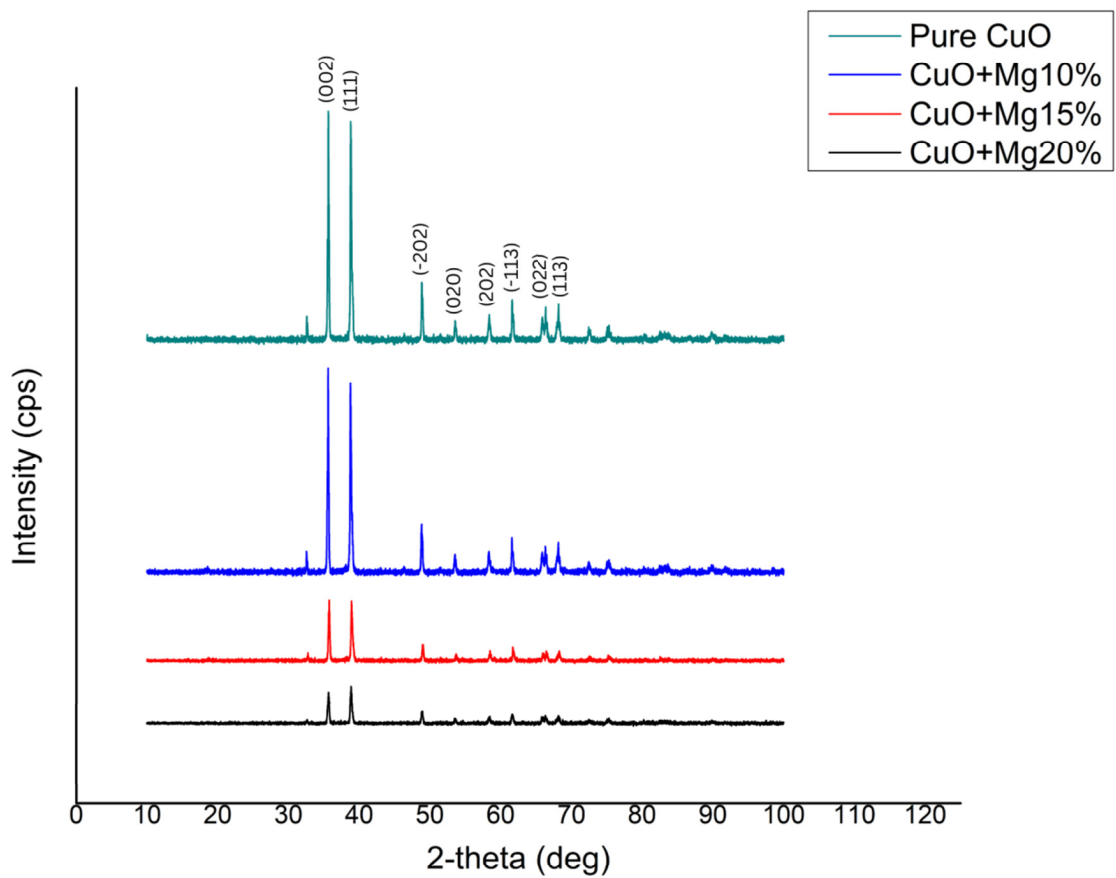

Figure 1: XRD pattern of Mg/CuO nanocomposites.

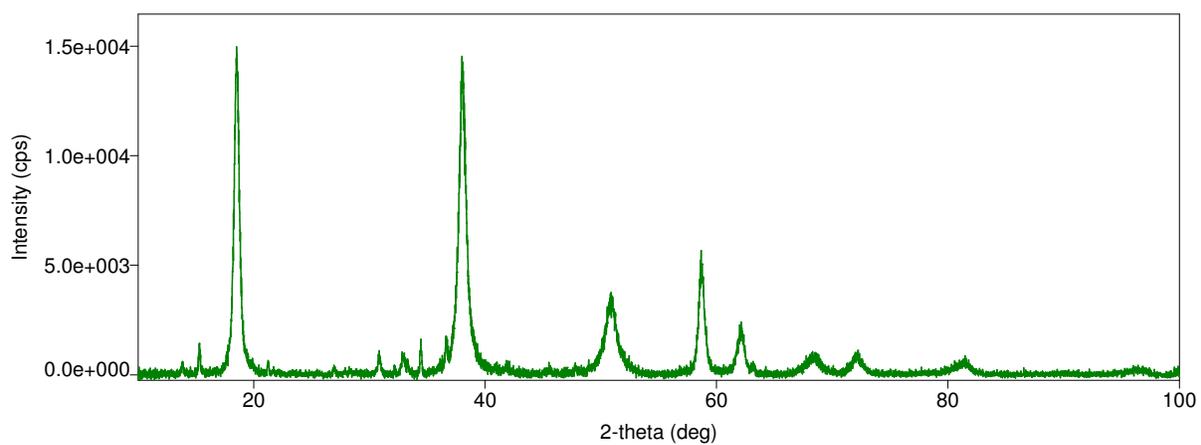

Figure 2: XRD pattern of pure Mg.

Crystallite size, dislocation density and strain of the samples are given in table 1. The crystallite size (*D*) was calculated using

$$D = \frac{0.91\lambda}{\beta \cos\theta} \quad (1)$$

Where $\lambda$ is the wavelength of Cu-K$\alpha$ radiation ($\lambda$=1.54060 $^0$A), and $\beta$ is the full width at half maximum (FWHM) of XRD peak in radians at angle $\theta$.

The dislocation density ($\delta$) was found using [21, 22, 23]

$$\delta = \frac{1}{D^2} \quad (2)$$

The strain ($\varepsilon$) was estimated using

$$\varepsilon = \frac{\beta \cos\theta}{4} \quad (3)$$

Table 1: Crystallite size, dislocation density and strain of samples.

| Mn Concentration | angle 2θ(deg) | angle θ (deg) | FWHM (deg) | Crystallite size (nm) | Dislocation density ($10^{14}$ lines/m$^2$) | Strain |
|---|---|---|---|---|---|---|
| Pure CuO | 35.62 | 17.81 | 0.17 | 49.61 | 4.06 | 0.000706472 |
| 10% | 35.58 | 17.79 | 0.16 | 52.71 | 3.60 | 0.00066499 |
| 15% | 35.74 | 17.87 | 0.18 | 46.87 | 4.55 | 0.000747777 |
| 20% | 38.84 | 19.42 | 0.25 | 34.06 | 8.62 | 0.001029134 |

(002) peak of XRD pattern given in figure 1 was used to calculate the values given in table 1. All the crystallites are in the range of nanometers. Filling dopant atoms in vacant lattice sites slightly increase the lattice size. As a result, the crystallite size initially increases. The crystallite size decreases at higher Mg concentrations. Due to the mismatch of the radius of foreign atom and local atoms, lattice deformation occurs. As a result, crystallite size decreases. On the other hand, adding foreign atoms donate nucleus to create new small particles. When the contribution of these new small particles in sample becomes significant, the average crystallite size decreases with the addition of new particles. The dislocation and strain become higher at higher Mg concentrations due to the deformation of the lattice.

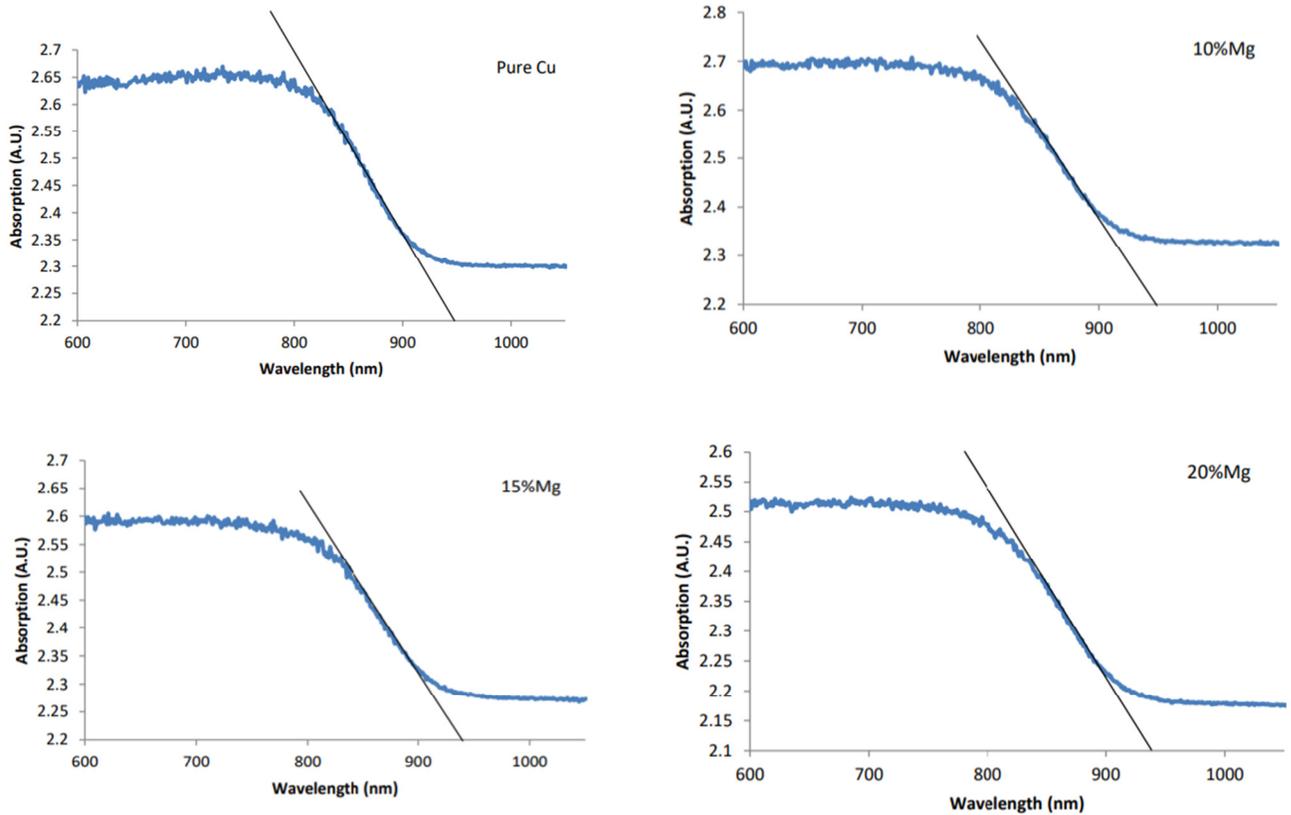

Figure 3: UV Visible spectrums of nanocomposite samples..

According to figure, the optical band gap does not vary significantly with the composition of the samples. Therefore, the variation of the gas sensitivity of composites is not attributed to the optical band gap. The calculated optical band gap is in the range of 1.31 to 1.32 eV, which are in the range of band gap of pure CuO. The optical band gap of metals such as Mg is negligible compared to the band gap of semiconductors such as CuO. As a result, the optical band gap of Mg/CuO composites is in the range of band gap of CuO. In addition, the highest absorption of all the samples does not vary significantly, since only CuO absorbs UV visible waves.

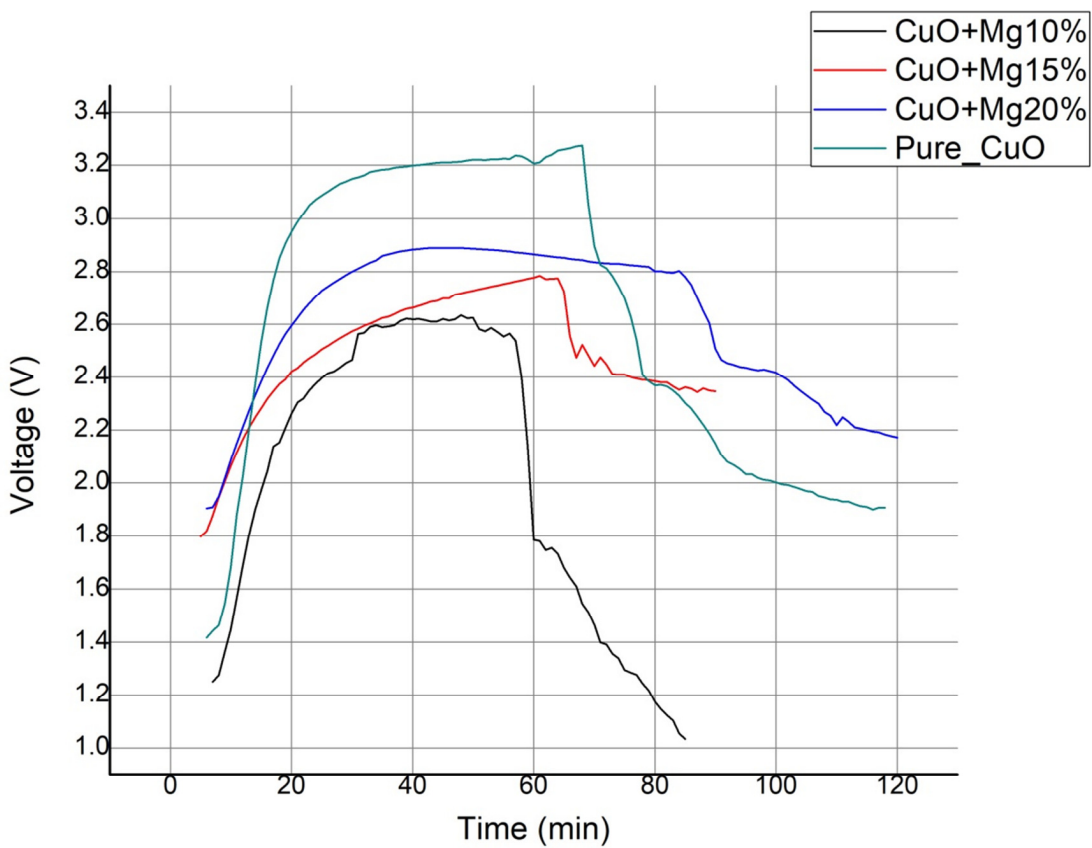

Figure 4: Voltage across standard resistor versus time.

Figure 4 indicates the graph of voltage measured across the standard resistor versus time. After adsorbing the methanol vapor, the voltage of the sample increases. While passing air through the chamber, the sample releases methanol vapor. As a result, the voltage decreases. Because the resistance of the sample varies with the amount of Mg, the maximum voltage of different samples are different.

The current of the sample was calculated from the following equation.

$V = IS$ → (4)

Where *V*, *I* and *S* are the voltage across the standard resistance shown in the figure 4, the current through the series circuit and the value of the standard resistor.

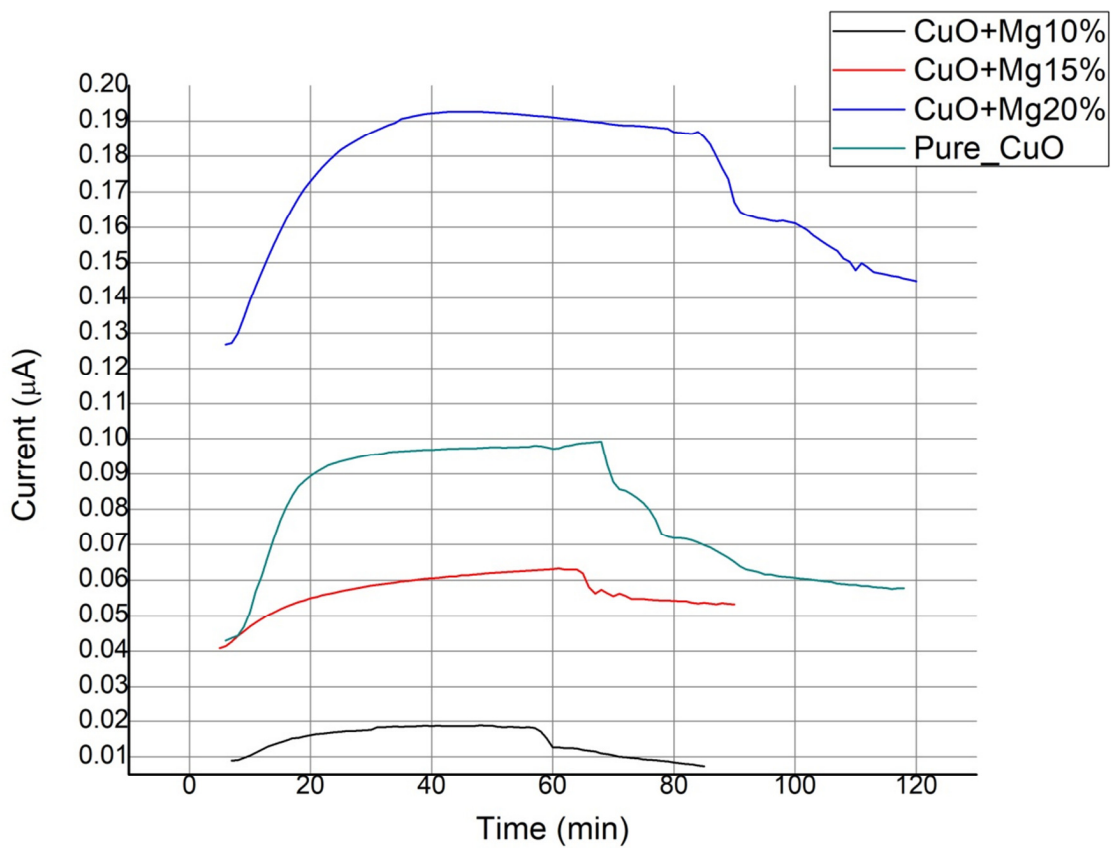

Figure 5: Electric current through the sample versus time.

Figure 5 represents the graph of current through the series circuit calculated from the equation number 4 versus time. Because the conductivity of the sample increases with the amount of Mg in the sample, the electrical current of the sample with 20%Mg is higher. In addition, because the area and the thickness of the samples slightly vary from the sample to sample, a slight variation of the initial resistance of the samples was noticed. Electric current initially increases due to the adsorption of methanol vapor. The current of the sample decreases in the process of releasing the methanol vapor.

The resistance of the sample was calculated from the following equation.

$5 = V+IR$ → (5)

Where $V$, $I$ and $R$ are the voltage across the standard resistance shown in the figure 4, the current through the circuit shown in figure 5 and the resistance of the sample.

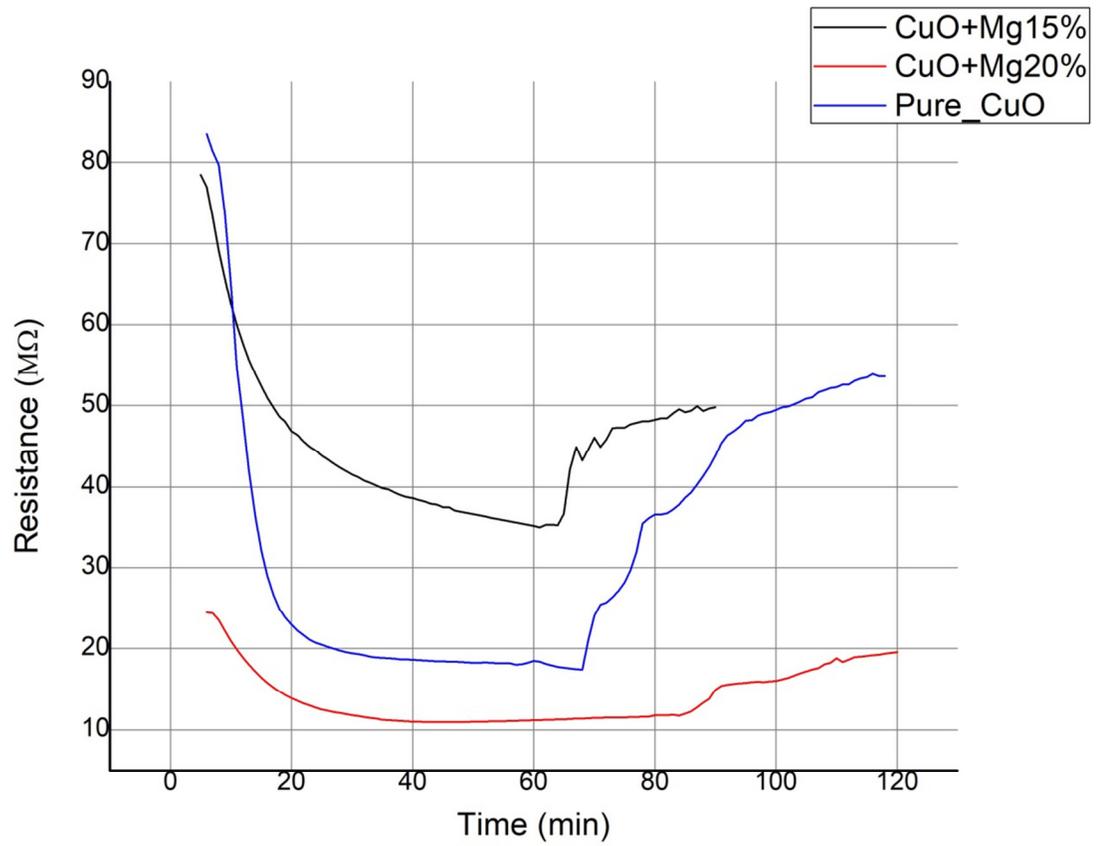

Figure 6: The resistance versus time for the pure Cu and samples with 15% and 20% Mg.

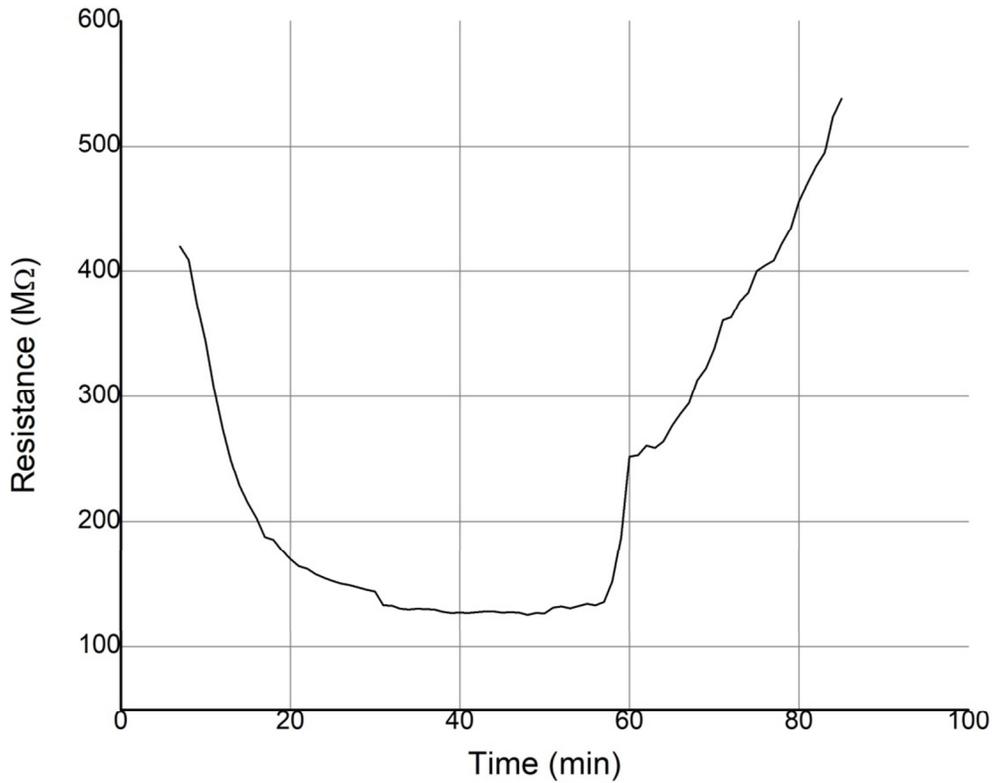

Figure 7: The resistance versus time for the sample with 10% Mg.

Figure 6 and 7 delineate the resistance of the sample found from equation number 5 versus time. The resistance of the sample decreases after adsorbing methanol vapor due to the reduction as given below. The depletion region of Mg/CuO samples decreases due to the adsorption of methanol vapor, resulting a decrease of the Schottky barrier.

V+O⁻ → VO+e⁻

Where V represents the methanol vapor. The donation of electrons to the sample increases the conductivity of the sample, resulting the decrease of the resistance.

The gas sensitivity ($S$) was found using the equation

$$S = \frac{R_i - R_f}{R_f} \rightarrow (6)$$

Where $R_i$ and $R_f$ are the initial resistance of the sample in air and the minimum resistance value of the sample in the gas, respectively.

Table 2: Gas sensitivity, response and recovery times of films in methanol vapor.

| Sample | Sensitivity | Response time (min) | Recovery time (min) |
|---|---|---|---|
| Pure CuO | 3.79 | 27 | 49 |
| CuO+10%Mg | 2.34 | 26 | 28 |
| CuO+15%Mg | 1.24 | 55 | 26 |
| CuO+20%Mg | 1.23 | 34 | 36 |

The best response and recovery times were found for the sample with 10% Mg. The gas sensitivity gradually increases with the CuO percentage in the nanocomposite. Although CuO atoms adsorb methanol vapor due to reduction reaction explained above, metals such as Mg do not adsorb any gas or vapor. Therefore, the gas sensitivity gradually increases with CuO content in composites. However, the best response time was measured for the sample with 10% Mg, and the best recovery time was found for the sample with 15% Mg.

**4. Conclusion:**

As per XRD analyses, the crystalline sizes of all the samples fall within the range from 34.06 to 52.71 nm, indicative of nanocomposite characteristics. The minimum crystallite size was observed at 20% Mg. The introduction of novel nucleus and lattice distortion attributable to Mg addition contributes to the observed variation in crystallite sizes. Optical band gaps range from 1.31 to 1.32 eV, signifying that Mg incorporation does not alter the band gap of CuO. The pure CuO exhibits the highest gas sensitivity, underscoring that changes in particle size or band gap are not determinants of gas sensitivity variation. Instead, it is the quantity of CuO within the sample that governs gas sensitivity. Prolonged response and recovery times are attributed to the absence of gold or silver catalytic under-layer during film preparation. At higher concentrations of Mg, the gas sensitivity did not change significantly when the Mg concentration was increased. When the contribution of Mg is dominant in the sample, the capability of CuO to detect the vapor is diminished.